\documentclass[sn-basic]{sn-jnl}
\usepackage{hyperref}%
\usepackage{fullpage}%

\usepackage{graphicx}%
\usepackage{multirow}%
\usepackage{amsmath,amssymb,amsfonts}%
\usepackage{amsthm}%
\usepackage{mathrsfs}%
\usepackage[title]{appendix}%
\usepackage{xcolor}%
\usepackage{textcomp}%
\usepackage{manyfoot}%
\usepackage{booktabs}%

\begin{document}

\title[Graded Double-Porosity Flow Solution]{Generalized Solution for Double-Porosity Flow through a Graded Excavation Damaged Zone}

\author[1]{\fnm{Kristopher} L. \sur{Kuhlman}}\email{klkuhlm@sandia.gov}

\affil[1]{\orgname{Sandia National Laboratories}, \orgdiv{Nuclear Waste Disposal Research \& Analysis}, \orgaddress{PO Box 5800, Mail Stop 0747}, \city{Albuquerque} \state{NM}, \postcode{87185}, \country{USA}}

\abstract{Prediction of flow to boreholes or excavations in fractured low-permeability rocks is important for resource extraction and disposal or sequestration activities. Analytical solutions for fluid pressure and flowrate, when available, are powerful, insightful, and efficient tools enabling parameter estimation and uncertainty quantification. A flexible porous media flow solution for arbitrary physical dimension is derived and extended to double porosity for converging radial flow when permeability and porosity decrease radially as a power law away from a borehole or opening. This distribution can arise from damage accumulation due to stress relief associated with drilling or mining.  The single-porosity graded conductivity solution was initially found for heat conduction, the arbitrary dimension flow solution comes from hydrology, and the solution with both arbitrary dimension and graded permeability distribution appeared in reservoir engineering. These existing solutions are here combined and extended to two implementations of the double-porosity conceptual model, for both a simpler thin-film mass transfer and more physically realistic diffusion between fracture and matrix. This work presents a new specified-flowrate solution with wellbore storage for the simpler double-porosity model, and a new more physically realistic solution for any wellbore boundary condition. A new closed-form expression is derived for the matrix diffusion solution (applicable to both homogeneous and graded problems), improving on previous infinite series expressions.}
  

\maketitle

\section{Introduction}
Fluid flow through damage-induced fracture networks in otherwise low-permeability crystalline rocks (e.g., granite, argillite or halite) is of interest to geothermal energy production  \citep{tao21}, radioactive waste disposal \citep{tsang05}, hydrogen storage \citep{abuaisha21}, and compressed air energy storage \citep{kim12}. Rock damage around an excavation (i.e., the Excavation Damaged Zone, EDZ; \citet{davies05}) increases the connected porosity, and leads to increased permeability. Fractured rock often has higher porosity and permeability than intact rock. Damage near a borehole or excavation will decrease the relative contribution from flow in the lower-permeability far-field, and will confound the estimation of hydrologic properties using approaches that assume uniform homogeneous distributions of permeability and porosity. There is a need for a flexible analytical solution for flow to a borehole or excavation in the presence of damage, that includes wellbore storage, double-porosity flow, and variable flow dimension.  This is most evident in a mechanically weak, low-permeability medium like salt, but should also apply to other low-permeability fractured rocks like granite or shale.

In salt, the far-field (i.e., undamaged) permeability is unmeasurably low \citep{beauheim02} due to salt's tendency to creep shut any unsupported openings. The permeability around a borehole in salt is derived from accumulated damage due to stress redistribution around the excavation itself \citep{wallace90,stormont91,cosenza96,hou03,kuhlman14}.

\begin{figure}[htb]
  \centering
  \includegraphics[height=0.4\textheight]{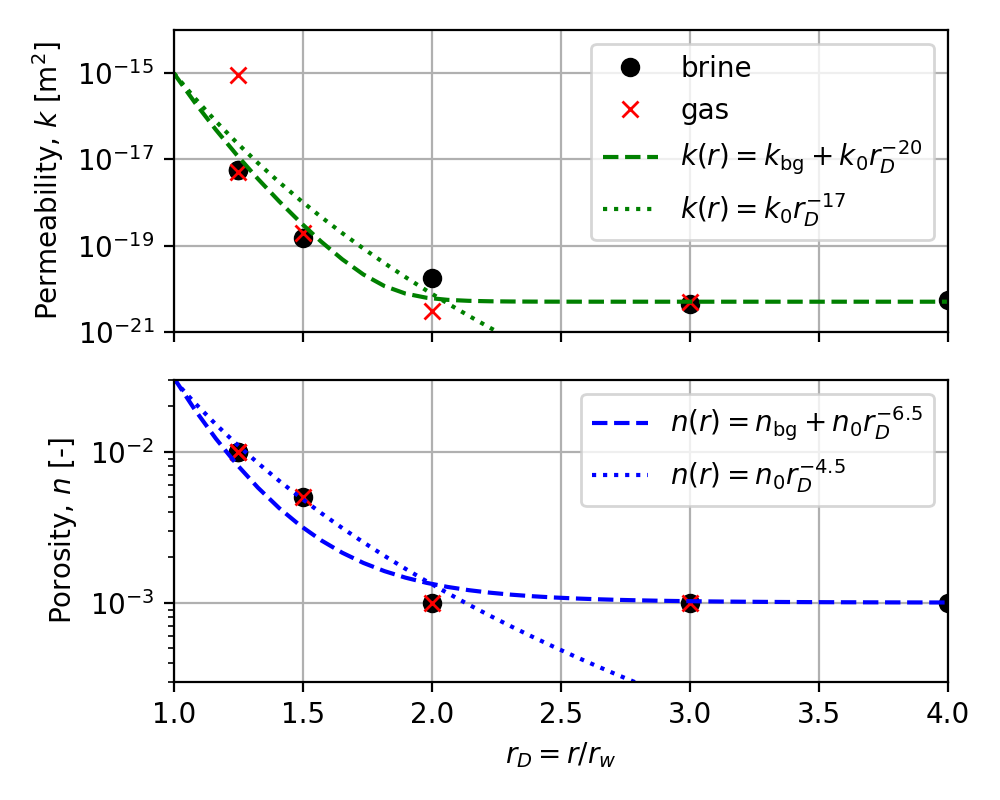}
  \caption{Permeability and porosity observations around a 1-m borehole (radial distance scaled by excavation radius) in salt from small-scale mine-by experiment (data from \citet{stormont91})}
  \label{fig:data}
\end{figure}

\citet{stormont91} presented brine and gas permeability data measured in salt for packer-isolated intervals of small boreholes before and after a central 1-meter diameter borehole was drilled (i.e., a mine-by experiment). Figure~\ref{fig:data} shows these data support the conceptual model of permeability and porosity decaying away from an excavation. \citet{cosenza96} proposed the power-law model for permeability and porosity plotted in the figure. These data show porosity and permeability decrease with distance from the central excavation. Two lines are shown with to the data; one is a monomial power-law, the other includes an additive background term. The two curves differ primarily away from the excavation ($r/r_w\ge3$), where larger uncertainties in estimated porosity and permeability exist, for three reasons. First, the access drift EDZ (test conducted in the floor of a 5-m wide room) is superimposed on the 1-m borehole EDZ. Second, the small-diameter (2.5-cm) measurement boreholes themselves each have a small EDZ overprinted on the 1-m borehole EDZ. Lastly, the apparent background permeability may represent the measurement limit of the packer system used (i.e., compliance of the packer inflation elements and working fluid). Especially in salt, the undisturbed background permeability is near zero, and is difficult to measure consistently in the field \citep{beauheim02}. The power-law distribution of permeability matches the more certain near-field permeability distribution, and is conceptually more elegant than a finite domain or a flow domain with piece-wise heterogeneous properties (i.e., a higher-permeability EDZ adjacent to lower-permeability intact rock).  Other investigations have also shown porosity and permeability decaying away with distance from an excavation in crystalline rocks \citep{shen11,cho13,Ghazvinian15} and sedimentary rocks \citep{perras10,perras16}.

Salt permeability has been related to both the confining and shear stresses \citep{reynolds60,lai71,stormont93,alkan09}. Confining stresses reduce fracture aperture and bulk permeability, while shear stresses are associated with increased bulk permeability. \citet{aydan93} present solutions for radial and tangential plane stress and strain (i.e., dilatation or a change in porosity) around a circular excavation. Strain is proportional to $r_D^{-2}$ or $r_D^{-3}$ (where $r_D$ is radial distance into the formation scaled by the excavation size), depending on whether the region is experiencing elastic (exponent $2$) or plastic (exponent $\approx 3$) deformation. These relationships illustrate a possible behavior of rock in the EDZ. The true extent of the EDZ depends on drilling or excavation method, borehole or tunnel geometry, state of stress, and rock mechanical properties \citep{hudson09}. Softer or weaker sedimentary rocks like argillite or halite typically have a larger EDZ than stiffer or stronger rocks like granite.

There are several well-known empirical power-law relationships between porosity and permeability in fractured or granular media \citep[e.g.,][]{kozeny27,carman37} and many studies have discussed their applicability \citep{david94,kuhlman18}.  Permeability in fractured rocks is more sensitive to small changes in porosity than granular rocks (i.e., fractured rocks have higher pore compressibility resulting in larger exponents in porosity-permeability relationships). 

Based on evidence from these observations, graded dimensionless porosity is assumed to follow
\begin{equation}
\label{eq:npower}
n({r})=n_0 \left(\frac{{r}}{r_w}\right)^{-\eta},
\end{equation}
where $r_w$ is the borehole or excavation radius [$\mathrm{m}$], $n_0=n(r_w)$ is maximum porosity at the borehole wall, and $\eta$ is a dimensionless exponent (see Table~\ref{tab:notation} for a list of physical variables and notation). Using the same form, the graded permeability can be represented with the form
\begin{equation}
\label{eq:kpower}
k({r})=k_0 \left(\frac{{r}}{r_w}\right)^{-\kappa},
\end{equation}
where $k_0=k(r_w)$ is the maximum permeability [$\mathrm{m^2}$] at the borehole wall and $\kappa$ is another dimensionless exponent. Based on lab measurements on fractured granite, the empirical relationship $\kappa \approx 3\eta$ has been proposed \citep{kranz79,david94}. The \citet{stormont91} salt data (Figure~\ref{fig:data}) support $\eta=4.5$ and $\kappa=17$, which shows a somewhat faster-decaying permeability ($\kappa=3.8 \eta$) than seen in granitic rocks.

The power-law permeability and porosity distribution conceptual model presented here is an alternative to flow models using wellbore skin \citep{streltsova88,pasandi08}, finite domain \citep{gelbard92,lin16}, or low-permeability non-Darcy flow with a threshold gradient \citep{liu14,Liu17}. These three conceptualizations all lead to reduced contributions of flow from the far field, but only borehole skin can account for observed distributions of higher porosity or permeability near the excavation, which are important when analyzing pressure or flowrate data at early time. The contribution from lower permeability in the far field are more important at late time. Finite domains and skin can have analytical flow solutions, but low-permeability non-Darcy flow does not typically lend itself to analytical solutions.

\citet{barker88} developed a generalized solution for converging flow to a borehole with variable non-integer dimension, $D$. This conceptualization has been used to characterize flow in fractured systems, where lower-dimension (i.e., $D<3$) results associated with discrete fractures are more common than higher dimension results \citep{beauheim04,le2004equivalent,bowman2013,ferroud2018insights}. \citet{doe1991} extended the solution of \citet{barker88} to the conceptualization where permeability varies with radial distance, through analogy with the heat conduction literature \citep{carslaw59}.

A single-porosity flow solution is derived here with power-law variable properties, like the approach of \citet{doe1991} (who did not present a derivation). The single-porosity solution is then readily extended to a double-porosity conceptualization, using first the approach of \citet{warren1963behavior} for thin-film mass transfer between fractures and matrix, then the more physically realistic matrix diffusion approach of \citet{kazemi1969pressure}. 

Double-porosity flow is a common and efficient conceptualization in fractured rocks \citep{aguilera80,golf-racht82,daprat90}. The medium is conceptualized as two communicating physically overlapping continua including fractures with high permeability (but little to no storage) and matrix or intact rock with significant storage (but little to no flow) \citep{barenblatt60a,barenblatt60b}. Many extensions to the basic double-porosity conceptual model exist, including multiple matrix or fracture porosities, and different assumptions about the geometry or underlying physics governing flow in the fractures or matrix \citep{chen1989transient,kuhlman21}. The \citet{warren1963behavior} solution simplifies the exchange between matrix and fractures to a mass-transfer thin-film approximation, leading to numerous analytical solutions \citep{aguilera80,chen1989transient}. It is commonly used for this reason, even though it is well-known that spatial pressure gradients in matrix blocks are important, as the matrix is low-permeability and would therefore be expected to experience steep, slow-changing gradients. A series representation of the \citet{kazemi1969pressure} solution is used here, an extension of the multirate mass transfer model to double-porosity flow \citep{kuhlman2015multiporosity}. The more physically correct (but more difficult to solve) solution can be represented by an infinite series of porosities, which can either represent an infinite number of Warren-Root type matrix porosities, or if the coefficients are chosen specifically, a single Kazemi-type matrix diffusion porosity. More recently, \citet{wang21darcian} has developed a semi-analytical solution for flow in a double-porosity  formation, for the case when non-Darcian flow is significant. \citet{moutsopoulos22} have provided analytical and semi-analytical solutions for two classical problems in flow of unconfined double-porosity aquifers, based on \citet{moutsopoulos21}. \citet{de-smedt22} presented an analytical solution for flow in double-porosity media for fractional flow dimensions, which is a generalization of \citet{de-smedt11}. \citet{hayek18} presented a semi-analytical solution for flow due to pumping a double-porosity aquifer via a constant-pressure boundary condition (without wellbore storage) where permeability varied as a power law.

The fractal reservoir flow problem \citep{chang90} is also analogous to the radially variable properties approach presented here, but the governing equations of the two problems are only equivalent when the spectral exponent ($\theta$ in \citet{chang90}) in the fractal problem is zero. The fractal reservoir governing equation is typically solved approximately, since the additional terms due to non-zero spectral exponent in the governing equation do not readily allow closed-form analytical solution.

In the next section, the governing equations and boundary conditions are developed for the variable-dimension single-porosity flow problem \citep{doe1991}. This solution is mapped onto the modified Bessel equation, allowing solution for flow to both specified pressure (type-I) and specified flowrate with wellbore storage (type-III). These more general single-porosity solutions are shown to degenerate down to several well-known cases. The single-porosity solutions are then extended to a simpler Warren-Root type double-porosity model for type-I \citep{hayek18} and type-III (new) and then a new Kazemi type double-porosity model. The Kazemi series solution approach is then summed analytically to arrive at a new closed-form expression for the response in Laplace space, a solution that is new for both graded and homogeneous domains. Finally, a summary and discussion of limitations is given for the new solutions. 

The approach taken here, representing the porosity and permeability of fractured rocks as power-law distributions, was first developed by \cite{delay07}, and first pursued by the author for applications in deep ($>3$ km) borehole disposal of radioactive waste in basement rock \citep{brady2017deep,kuhlman19}. The approach is also applicable to flow in salt surrounding excavations, like those in mine-by experiments \citep{stormont91}.

\section{Development of Flow Problem}
To introduce and contrast with the dual-porosity solution, the single-porosity solution is developed first. To make a single solution for Cartesian linear, cylindrical, and spherical geometries, a variable-dimension approach like \citet{barker88} is used, including variable permeability and porosity, like \citet{doe1991}. The governing equation for slightly compressible time-dependent change in pressure $p$ [$\mathrm{Pa}$] in a general 1D coordinate \citep{barker88} is
\begin{equation}
  \label{eq:gov}
  n(r) c \frac{\partial p}{\partial t} = \frac{1}{r^m} \frac{\partial}{\partial  r} \left[ \frac{k(r)  r^m}{\mu} \frac{\partial p}{\partial  r}\right],
\end{equation}
where $c$ is bulk compressibility $\mathrm{[1/Pa]}$ and the dimensionless parameter $m$ is 0 for a Cartesian strip, 1 for a cylinder, and 2 for a sphere (i.e., $m=D-1$, where $D$ is the dimension). The derivative of the bracketed term in \eqref{eq:gov} is expanded via chain rule; starting from \eqref{eq:kpower}, $\frac{\mathrm{d} k}{\mathrm{d} r}=-\kappa k(r)/r$ is substituted with the definitions of $k(r)$ and $n(r)$, to get 
\begin{equation}
  \label{eq:expand2}
  n_0c \left(\frac{{r}}{r_w}\right)^{-\eta} \frac{\partial p}{\partial t} = \frac{k_0}{\mu} \left(\frac{{r}}{r_w}\right)^{-\kappa} \left[ \frac{m-\kappa}{ r} \frac{\partial p}{\partial  r} + \frac{\partial^2 p}{\partial  r^2} \right].
\end{equation}

For converging radial flow in a semi-infinite domain, the relevant wellbore boundary conditions are constant-pressure (type-I), constant-flux (type-II), or constant-flux with wellbore storage (type-III in Laplace space). The initial, far-field, and source borehole boundary conditions for a borehole in an infinite symmetric domain are
\begin{align}
  \label{eq:BC}
  \mathrm{initial}\quad p({r},t=0) &=0 \nonumber\\
  \mathrm{far-field}\quad p({r} \rightarrow \infty,t) &< \infty \nonumber \\
  \mathrm{wellbore\;type-I}\quad p^{I}(r=r_w,t) &= p_1(t); \quad\mathrm{or} \\            
  \mathrm{wellbore\;type-II}\quad\left.\frac{A_m k_0}{\mu}\frac{\partial p^{II}(t)}{\partial r}\right|_{r=r_w}&=Q(t); \quad \mathrm{or} \nonumber \\
  \mathrm{wellbore\;type-III}\quad\frac{A_m k_0}{\mu} \left. \frac{\partial p^{III}(t)}{\partial r} \right|_{r=r_w} &= Q(t)+ \frac{A_c}{\rho g} \frac{\partial p_w(t)}{\partial t},\nonumber
\end{align}
respectively. See Appendix A for definition of source borehole boundary condition terms. These boundary conditions represent a homogeneous uniform initial condition, a requirement that the solution remains finite at large distance, and a specified pressure or pressure gradient at the source ($r=r_w$).

The Type-II boundary condition (specified flowrate) is a special case ($\sigma=0$) of the wellbore storage boundary condition (flowrate linearly proportional to change in pressure), so it is not developed further. 

\subsection{Dimensional Analysis}
A solution is derived for equation~\eqref{eq:expand2}, using the approach of \citet{doe1991}, which was based on analogy with the heat conduction literature \citep{carslaw59}. Reducing the governing equation \eqref{eq:expand2} to dimensionless form using characteristic time, $T_c=n_0 c L_c^2 \mu / k_0$, and characteristic length, $L_c=r_w$, leads to
\begin{equation}
  \label{eq:non-dim-gen}
  r_D^{\kappa-\eta} \frac{\partial p_D}{\partial t_D} = \frac{m-\kappa}{r_D}\frac{\partial p_D}{\partial r_D} + \frac{\partial^2 p_D}{\partial r_D^2},
\end{equation}
where the dimensionless quantities $r_D = r/L_c$, $t_D=t/T_c$, and $p^{\lbrace I,III \rbrace}_D=p/p^{\lbrace I,III \rbrace}_c$ are used (see Table~\ref{tab:dimless} for a summary of dimensionless quantities).

The characteristic pressure change is given by $p^{I}_c=\hat{p}_1$, where $p_1(t) = \hat{p}_1 f_t$ separates the time-dependent specified pressure into a constant characteristic pressure and a dimensionless variable time behavior (for a constant specified pressure, $f_t=1$). The dimensionless type-I initial and boundary conditions are
\begin{align}
  \label{eq:non-dim-s-BC}
  p_D(r_D,t_D=0)&=0 \nonumber \\
  p_D(r_D \rightarrow \infty,t_D)&<\infty \\
  p^{I}_D(r_D=1,t_D) &= f_t. \nonumber
\end{align}

Using $p^{III}_c=\frac{r_w \hat{Q} \mu}{A_m k_0}$, where $Q(t)=\hat{Q} f_t$ similarly separates the time-dependent volumetric flowrate into a constant characteristic flowrate and a dimensionless time behavior. The dimensionless type-III source borehole boundary condition is
\begin{equation}
  \label{eq:non-dim-WB-BC}
\left. \frac{\partial p^{III}_D}{\partial r_D} \right|_{r_D=1} = f_t + \sigma \frac{\partial p^{III}_D}{\partial t},
\end{equation}
where $\sigma$ is a dimensionless wellbore storage coefficient (see Appendix A) and the same initial and far-field conditions apply as the type-I case.

\subsection{Laplace Transform}
Taking the dimensionless Laplace transform $\left (\bar{f}(s)=\int_0^{\infty} e^{-s t_D} f(t_D) \;\mathrm{d}t_D \right)$ of the governing partial differential equation \eqref{eq:non-dim-gen} (without loss of generality assuming zero initial condition) leads to the ordinary differential equation 
\begin{equation}
  \label{eq:non-dim-lap-gen}
   \frac{\mathrm{d}^2 \bar{p}_D}{\mathrm{d} r_D^2} + \frac{m-\kappa}{r_D}\frac{\mathrm{d} \bar{p}_D}{\mathrm{d} r_D} - s\bar{p}_D r_D^{\kappa-\eta}= 0,
\end{equation}
assuming $\kappa$, $\eta$, and $m$ are not functions of time, and $s$ is the dimensionless Laplace transform parameter.

The transformed type-I and far-field boundary conditions \eqref{eq:non-dim-s-BC} are
\begin{align}
  \label{eq:non-dim-lap-s-BC}
  \bar{p}_D(r_D \rightarrow \infty)&<\infty \\
   \bar{p}^{I}_D(r_D=1)&=\bar{f}_t, \nonumber
\end{align}
where $\bar{f}_t$ represents the Laplace transform of the boundary condition's time behavior. For a unit step change at $t=0$ (where $f_t=1$, a typical assumption), $\bar{f}_t=\frac{1}{s}$. Other temporal behaviors are simply handled, including a step change at a non-zero time, an exponentially decaying source term, an arbitrary piecewise-constant or piecewise-linear behavior, or a sinusoidal source term \citep{kruseman94,mishra2013simulation}.

The transformed wellbore-storage boundary condition is
\begin{equation}
  \label{eq:non-dim-lap-WB-BC}
  \left. \frac{\mathrm{d} \bar{p}^{III}_D}{\mathrm{d} r_D} \right|_{r_D=1} = \bar{f}_t + \sigma s \bar{p}^{III}_D,
\end{equation}
which now more clearly resembles a Type-III boundary condition.

\subsection{Numerical Inverse Laplace Transform}
The governing equations and associated boundary conditions are solved exactly in Laplace space, then numerically inverted back to the time domain using one of several viable approaches \citep{kuhlman13}. The equations were rapidly prototyped and inverted using the Python library mpmath \citep{mpmath}, which provides arbitrary precision special functions and numerical inverse Laplace transform algorithms. A Fortran program was also developed to facilitate plotting and parameter estimation, implementing the inversion algorithm of \citet{dehoog1982}. Python and Fortran implementations of the solution are available at \url{https://github.com/klkuhlm/graded}.

\section{Solution of Flow Problem}
\subsection{Mapping onto Modified Bessel Equation}
The governing ordinary differential equation \eqref{eq:non-dim-lap-gen} can be made equivalent to a form of the modified Bessel equation after a change of variables first used by \citet{lommel68} for the standard Bessel equation. Appendix B illustrates an analogous change of variables to the modified Bessel equation. Comparing \eqref{eq:non-dim-lap-gen} to this scaled version of the modified Bessel equation \eqref{eq:app9}, they are equivalent given the following correspondences
\begin{align}
  \label{eq:lommel-equiv}
  \alpha=&\frac{1}{2} \left( \kappa - m + 1\right) & \gamma=&\frac{1}{2}\left( \kappa - \eta + 2\right) \\
  \nu=&\sqrt{\frac{\alpha^2}{\gamma^2}}=\frac{\kappa-m+1}{\kappa-\eta+2} & \beta&=\sqrt{\frac{s}{\gamma^2}}=\sqrt{\frac{4s}{\left( \kappa - \eta + 2\right)^2}}.  \nonumber
\end{align}  

The transformed modified Bessel equation has the general solution \eqref{eq:app1}
\begin{equation}
  \label{eq:lommel-soln}
    y = z^{\alpha} \left[ A \mathrm{I}_\nu \left( \beta z^\gamma \right) + B \mathrm{K}_\nu \left( \beta z^\gamma \right)\right], \qquad \left(\gamma \ne 0\right),
\end{equation}
where $A$ and $B$ are constants determined by the boundary conditions and $\mathrm{I}_\nu(z)$ and $\mathrm{K}_\nu(z)$ are the first- and second-kind modified Bessel functions of non-integer order and real argument \citep{mclachlan55,bowman58,spanier87,NIST:DLMF}. 

The finiteness boundary condition \eqref{eq:non-dim-lap-s-BC} requires $A=0$ to keep the solution finite as $r_{D} \rightarrow \infty$, since the first-kind modified Bessel function grows exponentially with increasing real argument, leaving
\begin{equation}
  \label{eq:gen-soln}
  \bar{p}_D \left( r_D \right) = r_D^{\alpha} B \mathrm{K}_{\nu} \left(\beta  r_D^{\gamma} \right),
\end{equation}
which is not defined for $\gamma=0$ (i.e., $\kappa-\eta=-2$, which is unrealistic because $\kappa$ is larger than $\eta$ for physical reasons), and $B$ is determined by the Laplace-space source borehole boundary conditions. 

\subsection{Constant-Pressure (Type-I) at Borehole}
The borehole boundary condition ($r_D=1$) for specified change in pressure leads to the solution (the \cite{warren1963behavior} double porosity solution for this wellbore boundary condition is equivalent to \citet{hayek18})
\begin{equation}
  \label{eq:soln-s-r}
  \bar{p}^{I}_D(r_D) =  \bar{f}_t r_D^{\alpha} \frac{  \mathrm{K}_\nu \left( \beta r_D^{\gamma} \right)}{ \mathrm{K}_\nu\left( \beta \right) }
\end{equation}
and its radial gradient (i.e., proportional to flow of fluid into the borehole)
\begin{equation}
  \label{eq:soln-grad-r}
  \frac{\mathrm{d} \bar{p}^{I}_D}{\mathrm{d} r_D} = \bar{f}_t r_D^{\alpha-1} \left [ \left( \alpha - \gamma \nu \right)  \frac{\mathrm{K}_\nu\left( \beta r_D^\gamma \right)}{\mathrm{K}_\nu\left( \beta \right)} + \beta \gamma r_D^\gamma \frac{\mathrm{K}_{\nu-1}\left( \beta r_D^\gamma \right)}{\mathrm{K}_\nu\left( \beta \right) } \right],
\end{equation}
using a recurrence relationship for the derivative of the Bessel function in terms of Bessel functions of adjacent orders \citep[\S 10.29.2]{NIST:DLMF}.

Restricting $\kappa \ge \eta$ (i.e., permeability decreases as fast or faster than porosity), then $\gamma > 0$ and $\alpha=\gamma \nu$ (for $\gamma<0$, $\alpha-\gamma\nu=2\alpha$). This physically motivated restriction on parameters simplifies \eqref{eq:soln-grad-r} to
\begin{equation}
  \label{eq:soln-grad-r-simp}
  \frac{\mathrm{d} \bar{p}^{I}_D}{\mathrm{d} r_D} = \sqrt{s} \bar{f}_t r_D^{\alpha+\gamma-1}  \frac{\mathrm{K}_{\nu-1}\left( \beta r_D^\gamma \right)}{\mathrm{K}_\nu\left( \beta \right) } ,
\end{equation}
since $\beta \gamma = \sqrt{s}$ for $\gamma>0$.  When evaluated in the source borehole ($r_D=1$), the solution simplifies further.

\begin{figure}
  \centering
  \includegraphics[width=\textwidth]{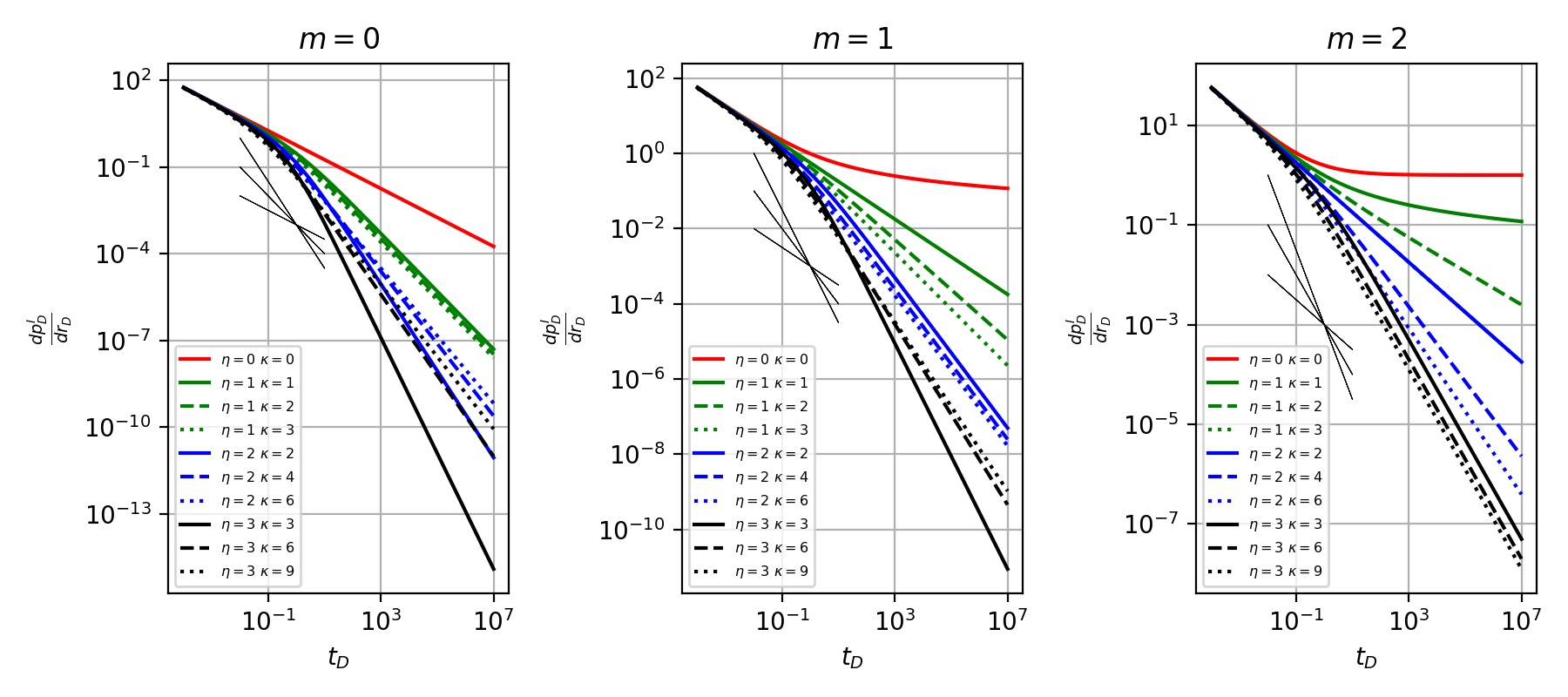}\\
  \includegraphics[width=\textwidth]{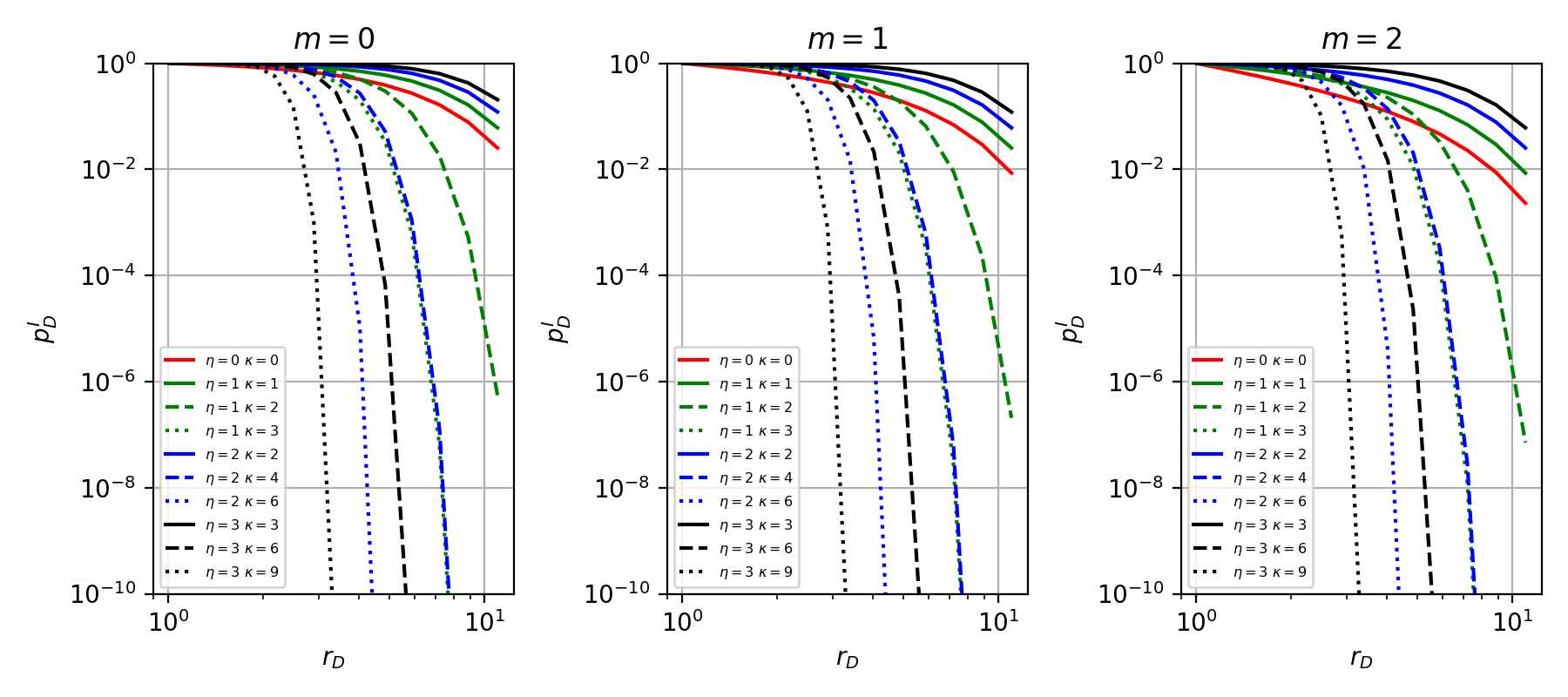}
  \caption{Type-I flowrate (top row at $r_D=1$) and pressure (bottom row at $r_D>1$ and $t_D=10$) solution at borehole for $m=0,1,2$ (Cartesian, cylindrical, and spherical) and at different radial distances. Line color indicates $\eta$; line type indicates $\kappa/\eta$. Line segments in top row illustrate slopes of $1/2$, $1$, and $3/2$.}
  \label{fig:typeI}
\end{figure}

Figure~\ref{fig:typeI} shows plots of the predicted pressure gradient at $r_D=1$ due to a constant-pressure condition there (top row) and the predicted decrease in pressure radially away from the boundary (values of $\eta$, $\kappa$, and $m$ for each simulation are listed in the caption and title of each figure). Both rows of plots show the variability with the porosity exponent ($\eta$, given by the line color) and the permeability exponent ($\kappa=\eta\tau$, given by the line type). The same results are shown for Cartesian linear ($m=0$), cylindrical ($m=1$), and spherical ($m=2$) geometries in three columns.

For a given set of parameters, a higher-dimensional domain (larger $m$) leads to a slower drop in produced fluids at any time. The highest sustained flowrate for all dimensions is achieved with constant properties in space (i.e., the red curve $\eta=\kappa=0$). More negative exponents in the porosity and permeability power-laws lead to more rapid decrease in flowrate, as the contribution to flow from large radius vanishes when the exponent increases in magnitude. These types of responses might be mis-interpreted as being associated with lower permeability (which would also lead to a faster decrease in flowrate) using a model with constant properties and a fixed dimension.

In the source well (top row of subplots), the effect of $\kappa$ is different and are predicted to reverse between dimensions. For $\eta=3$ (black lines), the $\kappa=\left\{ 3,6,9\right\}$ cases are swapped between $m=1$ and $m=2$. For $\eta=2$ (blue lines), the $\kappa$ cases are swapped between $m=0$ and $m=1$. 

The bottom row of figures shows the predicted pressure with distance at $t_D=10$. At locations away from the source well ($r_D>1$), changes in the porosity exponent, $\eta$, have relatively less impact than changes in the permeability exponent, $\kappa$ (different colored solid lines are close together, while colored lines of different line type are widely separated). The dimensionality ($m$) has a smaller effect at locations away from the source borehole than it had on the gradient predicted at the source borehole. 

\subsection{Constant-Flowrate with Wellbore Storage (Type-III)}
The wellbore-storage boundary condition for the specified flowrate solution at $r_D=1$ results in the general solution (that is new for any double-porosity solution with power-law variation in material properties)
\begin{equation}
  \label{eq:soln-WB-r}
  \bar{p}^{III}_D(r_D) =  \bar{f}_t r_D^{\alpha}\frac{ \mathrm{K}_\nu \left( \beta r_D^{\gamma} \right)}{  \left( \alpha -\gamma \nu +\sigma s\right)\mathrm{K}_{\nu}\left( \beta \right) + \beta \gamma \mathrm{K}_{\nu-1}\left( \beta \right)},
\end{equation}
which can be simplified using $\alpha=\gamma\nu$ and $\beta\gamma=\sqrt{s}$ to
\begin{equation}
  \label{eq:soln-WB-r-simp}
  \bar{p}^{III}_D(r_D) =  \bar{f}_t r_D^{\alpha} \frac{ \mathrm{K}_\nu \left( \beta r_D^{\gamma} \right)}{ \sqrt{s} \mathrm{K}_{\nu-1}\left( \beta \right) + \sigma s \mathrm{K}_{\nu} \left( \beta\right)}.
\end{equation}

\begin{figure}
  \centering
  \includegraphics[height=0.4\textheight]{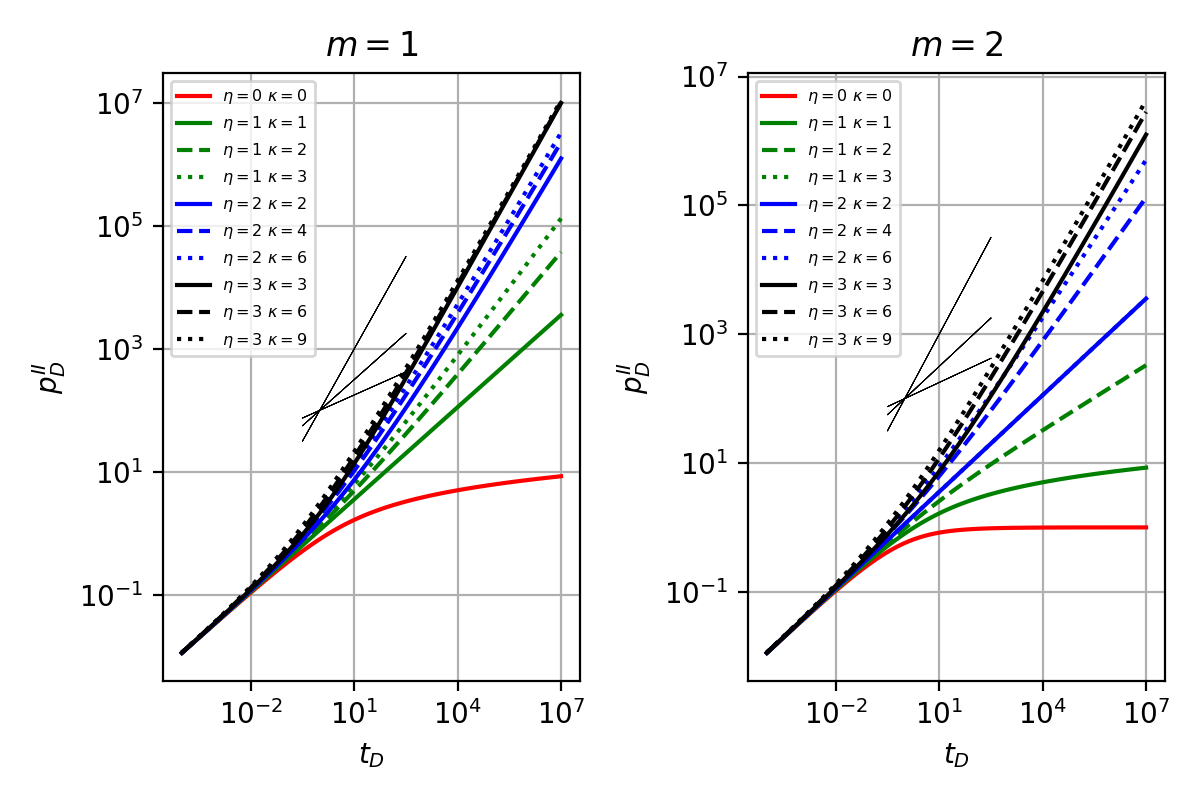}\\
  \caption{Type-II solution (Type-III with $\sigma=0$) at borehole for $m=1,2$ (cylindrical and spherical). Line color indicates $\eta$; line type indicates $\kappa/\eta$.}
  \label{fig:typeII}
\end{figure}

\begin{figure}
  \centering
  \includegraphics[height=0.4\textheight]{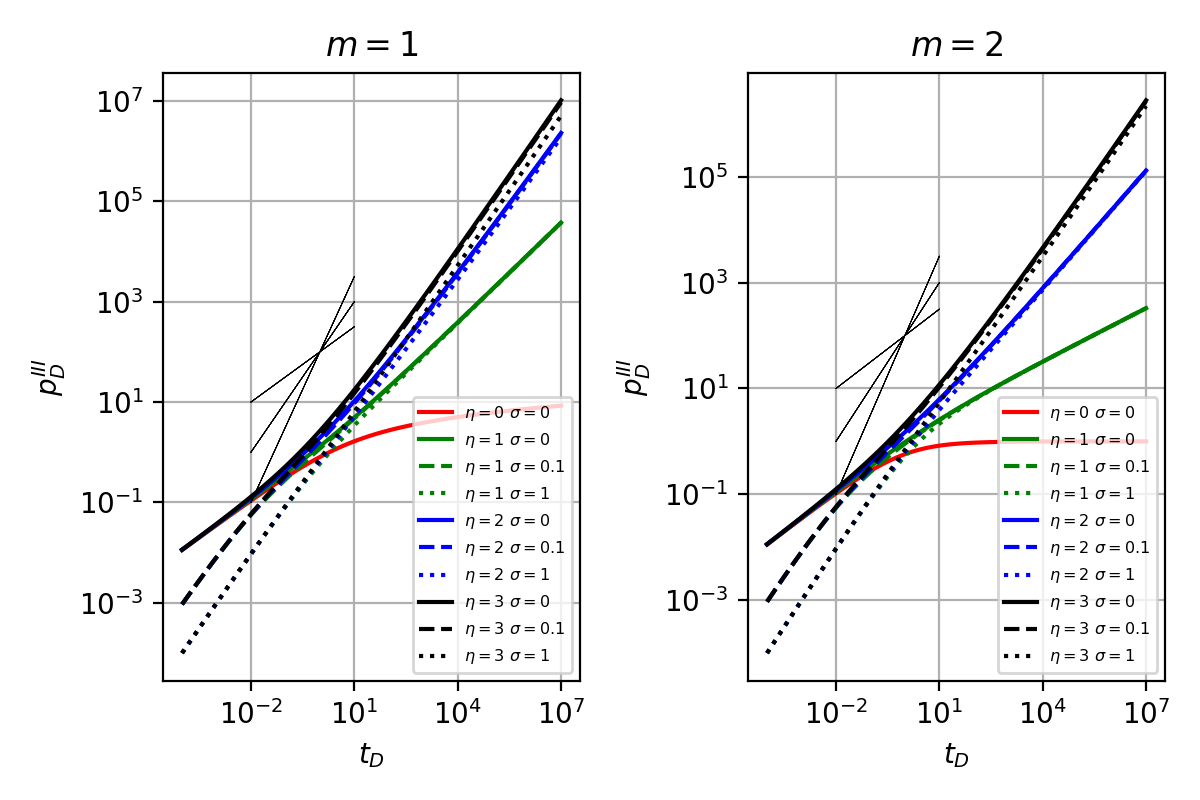}\\
  \caption{Type-III solution at borehole ($r_D=1$), for $m=1,2$ (cylindrical and spherical). Line color indicates $\eta$; line type indicates $\sigma$. All curves for $\kappa/\eta=2$.}
  \label{fig:typeIII}
\end{figure}

Analogous to the results for the Type-I solution but only showing the $m=1$ and $m=2$ cases, Figure~\ref{fig:typeII} shows the predicted pressure through time at the boundary for a specified flowrate at the boundary. Figure~\ref{fig:typeII} results are for no wellbore storage ($\sigma=0$), while Figure~\ref{fig:typeIII} shows the same results with non-zero wellbore storage (all model parameters listed in caption or title of each figure). Wellbore storage is important at early time, leading to a smaller predicted change in pressure, with the predicted response giving a characteristic $1:1$ slope on log-log plots before formation storage contributes significantly to the flow (i.e., pumping in a bathtub).  Wellbore storage makes more of a difference (i.e., shows a larger deviation from $\sigma=0$ case) for larger $\eta$ (and $\kappa$, since $\kappa=2\eta$). 

\subsection{Parameter Combinations Yielding Simpler Solutions}
When $\eta=\kappa=0$, permeability and porosity are constant in space; in this case \eqref{eq:non-dim-lap-gen} simplifies to
\begin{equation}
  \label{eq:eta-kappa-zeroa}
   \frac{\mathrm{d}^2 \bar{p}_D}{\mathrm{d} r_D^2} + \frac{m}{r_D}\frac{\mathrm{d} \bar{p}_D}{\mathrm{d} r_D} - s\bar{p}_D = 0,
\end{equation}
which is the dimensionless form of the equation solved by \citet{barker88}. In this case $\gamma=1$, $\alpha=(1-m)/2$, $\nu=\alpha$, and $\beta=\sqrt{s}$. The solution in Laplace-space under these conditions becomes
\begin{equation}
  \label{eq:eta-kappa-zerob}
  \bar{p}_D \left( r_D \right) = r_D^\nu B \mathrm{K}_\nu \left( \sqrt{s} r_D \right),
\end{equation}
which was found by \citet[Eqn.\ 15]{barker88}.

When $\eta=\kappa=m=0$ the time-domain solution simplifies to $p_D(t)=1/\sqrt{\pi t}$, because $\nu=1/2$ and $\nu-1=-1/2$, the numerator and denominator of \eqref{eq:soln-grad-r-simp} are equal since $\mathrm{K}_\nu(z)\equiv\mathrm{K}_{-\nu}(z)$. 

Another simplification occurs when $m=\kappa=\eta$, not necessarily zero. In this case, the permeability and porosity decrease at the same rate radially that the surface area of the domain grows in size ($A_0 \propto 1$, $A_1\propto r_D$, $A_2 \propto r_D^2$), resulting in an equivalent Cartesian coordinate system,
\begin{equation}
  \label{eq:eta-kappa-m}
   \frac{\mathrm{d}^2 \bar{p}_D}{\mathrm{d} r_D^2}  - s\bar{p}_D = 0,
\end{equation}
which has a solution in terms of $\sin(\sqrt{s}r_D)$ and $\cos(\sqrt{s}r_D)$ or $\exp(\pm \sqrt{s}r_D)$ and typically has an explicit inverse Laplace transform. In this case $\alpha=\nu=1/2$, $\gamma=0$, and $\beta=\sqrt{s}$.

When $\nu=n\pm \frac{1}{2}$ (for $n$ integer), the modified Bessel functions become modified spherical Bessel functions \citep[\S 10.47]{NIST:DLMF}, and when  $\nu=\pm \frac{1}{3}$, they become Airy functions \citep[\S 9.6]{NIST:DLMF}. These additional special cases are not handled differently here (i.e., the more general solution in terms of modified Bessel functions is still valid), since in the case given here $\nu$ varies with $\kappa$, $\eta$, and $m$ \eqref{eq:lommel-equiv}.

\section{Extension of Solution to Double Porosity}
\subsection{Mass-Transfer Coefficient Approximation}
Beginning with the \citet{warren1963behavior} formulation for double-porosity (i.e., high-conductance fractures and high-capacity matrix), the power-law permeability and porosity distributions are incorporated. The equations for double-porosity flow in the fractures and matrix are
\begin{align}
  \label{eq:wr-1}
   \frac{1}{r^m}\frac{\partial }{\partial r}\left[ \frac{k_f}{\mu} \frac{\partial p_f}{\partial r}\right]&= n_r c_r \frac{\partial p_r}{\partial t} + n_f c_f \frac{\partial p_f}{\partial t} \nonumber \\
\frac{\hat{\alpha} k_r}{\mu} \left( p_f - p_r\right)&=  n_r c_r  \frac{\partial p_r}{\partial t} 
\end{align}
where $\hat{\alpha}$ is the shape factor $\mathrm{[1/m^2]}$ of \citet{warren1963behavior}, subscript $f$ indicates fracture, and subscript $r$ indicates matrix (rock). The matrix equation does not involve a spatial gradient of pressure, nor a matching of pressure and flux at the boundary, but simply a difference between the fracture and matrix pressure (i.e., the mass transfer coefficient approximation often used for heat transfer across thin films). This behavior is sometimes referred to in the petroleum engineering literature as ``steady-state'' flow between the fracture and matrix  \citep{daprat90}, but it also represents one-dimensional diffusion in the matrix with a thin-film mass-transfer approximation between the fracture and matrix reservoirs, analogous to Newton's law of cooling. 

Substituting the permeability $k_i=k_{i0}\left(\frac{r}{r_w}\right)^{-\kappa_i}$ and porosity $n_i=n_{i0}\left(\frac{r}{r_w}\right)^{-\eta_i}$ ($i \in \{f,r\}$), then converting to dimensionless form using an analogous approach to \citet{warren1963behavior}, where $\omega=n_{f0}c_f/\left(n_{r0}c_r + n_{f0}c_f\right)$ is the dimensionless fracture storage coefficient and $\lambda=\hat{\alpha} k_r r_w^2/k_f$ is the dimensionless interporosity exchange coefficient. Finally, taking the Laplace transform of both equations results in the pair of ordinary differential equations
\begin{align}
  \label{eq:wr-4}
    \left[ \frac{\mathrm{d}^2 \bar{p}_{fD}}{\mathrm{d} r_D^2} + \frac{m - \kappa_f}{r_D} \frac{\mathrm{d} \bar{p}_{fD}}{\mathrm{d} r_D}\right] r^{-\kappa_f} &= (1-\omega)r_D^{-\eta_r} \bar{p}_{mD}s + \omega r_D^{-\eta_f}\bar{p}_{fD}s \nonumber \\
\lambda \left( \bar{p}_{fD} - \bar{p}_{rD}\right)r_D^{-\kappa_r} &=  (1-\omega) r_D^{-\eta_r}\bar{p}_{rD}s .
\end{align}
Solving for matrix pressure in the matrix equation, $\bar{p}_{rD}=\bar{p}_{fD}\lambda r_D^{-\kappa_r}/\left[ (1-\omega)s r_D^{-\eta_r} + \lambda r_D^{-\kappa_r}\right]$, and substituting this into the fracture equation leads to a single equation solely in terms of dimensionless Laplace-domain fracture pressure
\begin{align}
  \label{eq:wr-5}
    \left[ \frac{\mathrm{d}^2 \bar{p}_{fD}}{\mathrm{d} r_D^2} + \frac{m - \kappa_f}{r_D} \frac{\mathrm{d} \bar{p}_{fD}}{\mathrm{d} r_D}\right] r^{-\kappa_f} &= r_D^{-\eta_r} \bar{p}_{fD}  \left\lbrace\frac{(1-\omega) s r_D^{-\kappa_r}\lambda}{(1-\omega)s r_D^{-\eta_r} + \lambda r_D^{-\kappa_r}}\right\rbrace  + \omega r_D^{-\eta_f}\bar{p}_{fD}s .
\end{align}

To force the term in curly brackets in \eqref{eq:wr-5} to be independent of $r_D$, $\kappa_r=\eta_r$ is assumed. Setting $\kappa_r$ and $\eta_r$ equal to $\eta_f$ allows $r_D$ and $\bar{p}_{fD}$ to be similar form to previous solutions. Simplifying the subsequent notation $\kappa_f \rightarrow \kappa$, $\eta_r \rightarrow \eta$, and $\bar{p}_{fD} \rightarrow \bar{p}_{D}$ results in 
\begin{align}
  \label{eq:wr-6}
    \frac{\mathrm{d}^2 \bar{p}_{D}}{\mathrm{d} r_D^2} + \frac{m - \kappa}{r_D} \frac{\mathrm{d} \bar{p}_{D}}{\mathrm{d} r_D} &= r_D^{\kappa-\eta} \bar{p}_{D}  \left[\frac{(1-\omega) s \lambda}{(1-\omega)s + \lambda }  + \omega s \right], 
\end{align}
which is the same form as \eqref{eq:non-dim-lap-gen}. This solution corresponds to the same scaled Bessel equation, with only the definition of $\beta$ changing to
\begin{equation}
  \label{eq:wr-7}
  \beta_{WR} = \sqrt{\left[\frac{\lambda }{ \lambda/(1-\omega) + s}  + \omega \right] \frac{s}{ \gamma^2}}.
\end{equation}
Any more general spatial behavior of matrix properties (e.g., $\eta_r \ne \kappa_r$) would not be solvable with the same approach. This limitation still makes physical sense, as the the most important terms to vary with space are the fracture permeability and the matrix storage. Setting $\kappa=\eta=0$ and $m=1$ results in the  \citet{warren1963behavior} solution.

\begin{figure}
  \centering
  \includegraphics[height=0.4\textheight]{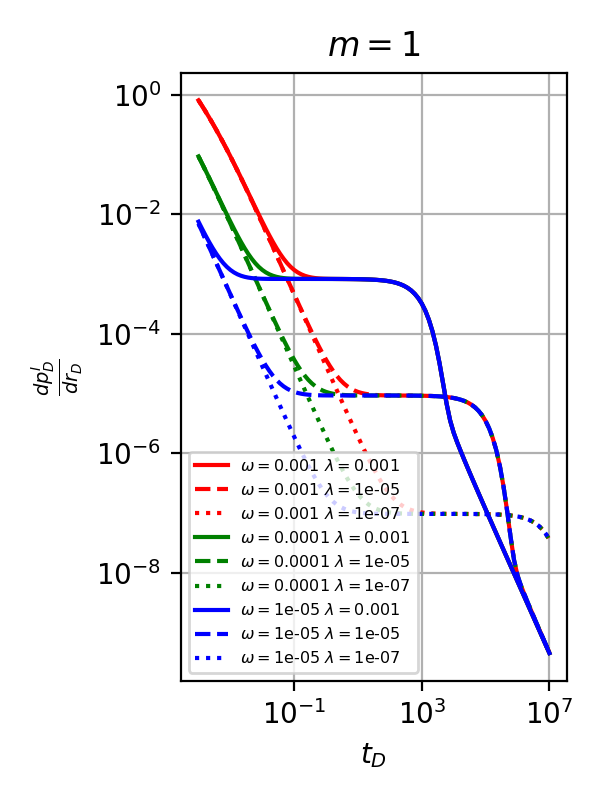}
  \includegraphics[height=0.4\textheight]{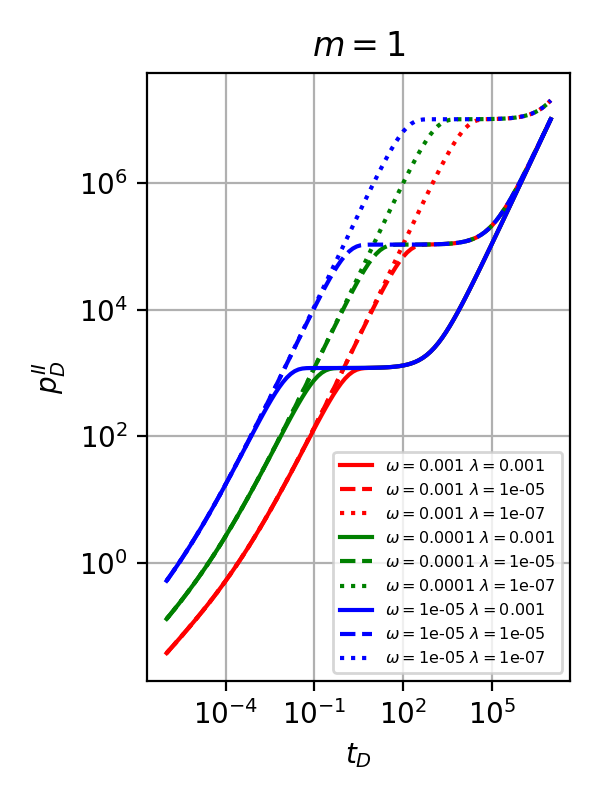}
  \caption{Type-I flowrate solution at borehole (left) and Type-II solution for pressure ($\sigma=0$, right), for $m=1$ (cylindrical). Line color indicates $\lambda$; line type indicates $\omega$.}
  \label{fig:WRa}
\end{figure}

\begin{figure}
  \centering
  \includegraphics[height=0.4\textheight]{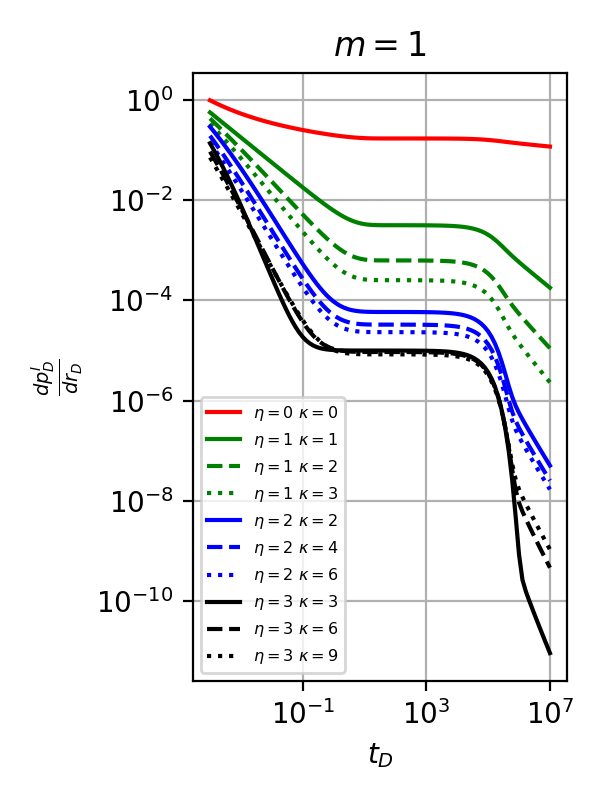}
  \includegraphics[height=0.4\textheight]{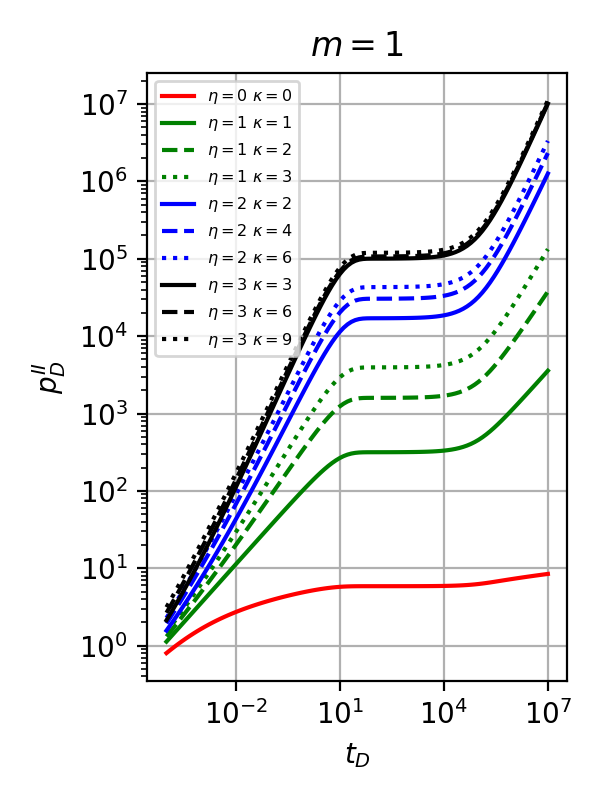}
  \caption{Type-I flowrate solution at borehole (left) and Type-II solution for pressure ($\sigma=0$, right), for $m=1$ (cylindrical). All curves are for $\lambda=10^{-5}$ and $\omega=10^{-4}$ (middle curves shown in Figure~\ref{fig:WRa}). Line color indicates $\eta$; line type indicates $\kappa/\eta$.}
  \label{fig:WRapl}
\end{figure}

Figure~\ref{fig:WRa} shows typical solution behaviors for the cylindrical ($m=1$) case for Type-I and Type-II wellbore boundary conditions, for $\eta=3$ and $\kappa=6$. Figure~\ref{fig:WRapl} shows behavior from the ``middle'' curve in Figure~\ref{fig:WRa} ($\lambda=10^{-5}$ and $\omega=10^{-4}$), for a range of porosity and permeability exponents similar to those shown in \cite{warren1963behavior}, listed in the figure caption. 

\subsection{Matrix Diffusion}
The matrix diffusion problem of \citet{kazemi1969pressure} is more physically realistic \citep{aguilera80,daprat90}, but it is typically solved numerically or via late-time approximations \citep{deswaan76}, rather than analytically like \citet{warren1963behavior}. The series approach of \citet{kuhlman2015multiporosity} is used here to represent matrix diffusion in a single matrix continuum through the sum of an infinite series of Warren-Root matrix continua, and the infinite sum is then analytically summed.

The generalization of \eqref{eq:wr-1} to multiple matrix continua starts with
\begin{align}
  \label{eq:MP-1}
   \frac{1}{r^m}\frac{\partial }{\partial r}\left[ \frac{k_f}{\mu} \frac{\partial p_f}{\partial r}\right]&= \sum_{j=1}^{N} n_j c_j \frac{\partial p_j}{\partial t} + n_f c_f \frac{\partial p_f}{\partial t} \nonumber \\
\frac{\hat{\alpha_j} k_j}{\mu} \left( p_f - p_j\right)&=  n_j c_j  \frac{\partial p_j}{\partial t} \qquad j=1,2, \dots N,
\end{align}
where $N$ is the number of matrix continua (one additional equation for each continuum). Similarly taking the Laplace transform of this set of equations, solving for $\bar{p}_f$, substituting the matrix equations into the fracture equation, and simplifying the notation leads to 
\begin{align}
  \label{eq:kz}
    \frac{\mathrm{d}^2 \bar{p}_{D}}{\mathrm{d} r_D^2} + \frac{m - \kappa}{r_D} \frac{\mathrm{d} \bar{p}_{D}}{\mathrm{d} r_D} &= r_D^{\kappa-\eta} \bar{p}_{D}  \omega s (1 + \bar{g}),
\end{align}
where
\begin{equation}
  \label{eq:kernel}
  \bar{g}=\sum_{j=1}^N\frac{\hat{\xi}_ju_j}{s + u_j}
\end{equation}
is a matrix memory kernel \citep{haggerty1995}, $\hat{\xi}_j$ is related to the storage properties of each matrix continuum (analogous to $\omega$ of \citet{warren1963behavior}), and $u_j$ is related to the interporosity flow coefficient of each matrix continuum (analogous to $\lambda$ of \citet{warren1963behavior}). The Laplace-space memory kernel approach is flexible, and is used elsewhere in hydrology and reservoir engineering \citep{herrera77,haggerty2000,schumer2003}. Equation \eqref{eq:kz} can be simplified to \citet{warren1963behavior} with a particular choice of $\bar{g}$ and $N=1$, and to the solution for a triple-porosity reservoir \citep{clossman75} with a different choice of $\bar{g}$ and $N=2$ \citep{kuhlman2015multiporosity}.

When  $N \rightarrow \infty$ in \eqref{eq:kernel}, the it is more convenient to specify the mean and variance of the parameter distributions than the individual parameters associated with each porosity. Several different distributions are possible \citep{haggerty1995}. In the form presented by \cite{kuhlman2015multiporosity}, the parameters are specified as the infinite series 
\begin{equation}
  \label{eq:kazemi-terms}
  u_j = \frac{(2j-1)^2 \pi^2 \lambda}{ 4 (1-\omega) } \qquad
  \hat{\xi}_j =\frac{8 (1-\omega)}{ (2j-1)^2 \omega \pi^2} \qquad j=1,2, \dots N \rightarrow \infty
\end{equation}
which leads to the \citet{kazemi1969pressure} solution for matrix diffusion. The parameters $\lambda$ and $\omega$ have the same definitions as in \citet{warren1963behavior}. 

Setting $\kappa=\eta=0$ results in the solution of \citet{kuhlman2015multiporosity}. The new governing equation is the same form and the modified Bessel function solution, only requiring re-definition of $\beta$ as
\begin{align}
  \label{eq:kazemi-beta}
  \beta_{KZ} = \sqrt{\left[ \sum_{j=1}^{N} \frac{\omega \hat{\xi}_j u_j}{u_j + s} + \omega \right] \frac{s}{\gamma^2}}, \qquad N \rightarrow \infty.
\end{align}
Substituting the definitions of $u$ and $\hat{\xi}$ from \eqref{eq:kazemi-terms} and simplifying leads to
\begin{align}
  \label{eq:kazemi-beta-simplified}
  \beta_{KZ} = \sqrt{\left[ \sum_{j=1}^{N} \frac{2 \lambda}{W^2_j \lambda/(1-\omega) + s} + \omega \right] \frac{s}{\gamma^2}}, \qquad N \rightarrow \infty,
\end{align}
where $W_j=\pi(2j-1)/2$. This is similar in form to \eqref{eq:wr-7} but the term in the denominator grows as the index increases, illustrating how the series solution approximates the \citet{kazemi1969pressure} solution through an infinite series of modified \citet{warren1963behavior} matrix porosities.

Further simplifying the approach of \citet{kuhlman2015multiporosity}, the infinite series in \eqref{eq:kazemi-beta-simplified} can be evaluated in closed form using residue methods \citep{Mathematica}, resulting in
\begin{equation}
  \label{eq:kazemi-beta-summed}
  \beta_{KZ} = \sqrt{\left[  \sqrt{\frac{ \lambda (1-\omega)}{s}} \tanh \left( \sqrt{\frac{s (1-\omega)}{\lambda}} \right) + \omega \right] \frac{s}{\gamma^2}},
\end{equation}
where $\tanh(\cdot)$ is the hyperbolic tangent. This closed-form expression derived here is more accurate and numerically more efficient than truncating or accelerating the infinite series in \eqref{eq:kazemi-beta}, which is an improvement over the series presented in \citet{kuhlman2015multiporosity} for graded or homogeneous domains.

\begin{figure}
  \centering
  \includegraphics[height=0.5\textheight]{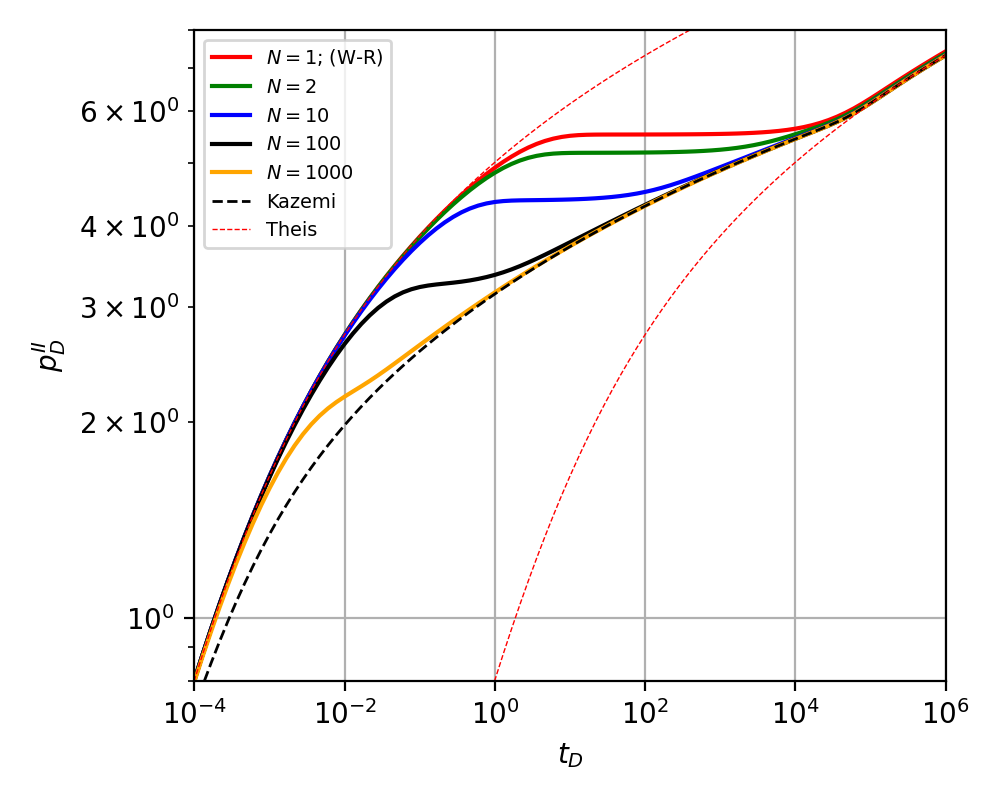}\\
  \caption{Type-II solution for pressure at source borehole ($\sigma=0$), for $m=1$ (cylindrical) for different number of terms. All curves are for $\lambda=10^{-5}$, $\omega=10^{-4}$, $\kappa=\eta=0$.}
  \label{fig:WRtoKZ}
\end{figure}

Figure~\ref{fig:WRtoKZ} illustrates the transition from the \citet{warren1963behavior} ($N=1$) to the \citet{kazemi1969pressure} series approximation for increasing terms ($N = \left\{ 2,10,100,1000\right\}$, heavy colored solid lines) and the expression for the infinite sum \eqref{eq:kazemi-beta-summed} (heavy black dashed line) for flow to a specified flux (type-II, $\sigma=0$) cylindrical ($m=1$) borehole of constant material properties ($\kappa=\eta=0$). The bounding \citet{theis1935} behavior is shown for the fracture and matrix compressibilities (thin red dashed lines).

\section{Applications and Limitations}
A general converging radial flow solution for specified flowrate or specified wellhead pressure was derived for domains with power-law variability in porosity and permeability due to damage. The single-porosity version has already been presented by \citet{doe1991}, and a solution for constant-pressure condition without wellbore storage was derived by \cite{hayek18}, but the specified-flowrate double-porosity solution with wellbore storage presented here is new. The infinite series approximation to Kazemi was summed analytically, resulting in a new closed-form expression of the series presented in \citet{kuhlman2015multiporosity}, which is an improvement for both graded and homogeneous properties. The newly developed analytical solutions are more general (i.e., several existing solutions are special cases of the new solution) and include more behaviors typical in well-test solutions (i.e., wellbore storage, positive skin, double porosity), while still being straightforward and parsimonious (i.e., as few free parameters as possible) in their implementation.

The basic flow solution assumes linear single-phase flow of a fluid in a slightly compressible formation. The double-porosity solution assumes the fractures are high permeability, with low storage capacity, while the matrix (i.e., intact rock between fractures) is high storage capacity with low permeability. These assumptions are representative for analytical solutions to subsurface porous media flow problems in the hydrology and petroleum engineering literature, and are shared by the solutions of \cite{barker88}, \cite{doe1991}, \cite{warren1963behavior}, \cite{kazemi1969pressure}, and \cite{kuhlman2015multiporosity}.

To apply this analytical solution to observed data, either observed data would be transformed into dimensionless space, or the analytical solution could be transformed to dimensional space, then a parameter estimation routine would be used to minimize the model-data misfit, and possibly explore the uncertainty or uniqueness of the solution. The solution method developed to solve these solutions uses numerical inverse Laplace transforms and runs quickly enough to be used in parameter estimation (e.g., Monte Carlo methods that require hundreds of thousands of evaluations).

The analytical solution might be of most use with parameter estimation to fit observations, but the non-uniqueness of the curves may make estimation of unique physical parameters difficult, without further physical or site-specific constraints. Realistically, the parameters in the Bessel equation may be estimable (i.e., $\alpha$, $\beta$, $\gamma$, and $\nu$ defined in \eqref{eq:lommel-equiv}), but without defining the flow dimension ($m$) or the relationship between the porosity and permeability exponents ($\tau=\kappa/\eta$), it may be difficult to identify all the parameters from data alone, since many the curves have similar shapes, unlike classical Type curves \citep{bourdet89}.

\begin{table}[htb]
  \centering
  \begin{tabular}{lll}
    \hline
    $A_c$ & borehole cross-sectional area & $\mathrm{m^2}$ \\
    $A_m$ & borehole cylindrical surface area & $\mathrm{m^2}$ \\
    $c$ & bulk compressibility & $\mathrm{1/Pa}$ \\
    $f_t$ & time variability & $\mathrm{-}$ \\
    $g$ & gravitational acceleration & $\mathrm{m/s^2}$ \\
    $h$ & hydraulic head & $\mathrm{m}$ \\
    $k$ & permeability & $\mathrm{m^2}$\\
    $L_c$ & characteristic length $(r_w)$ & $\mathrm{m}$ \\
    $m$ & dimension $(D-1)$ & $\mathrm{-}$\\
    $n$ & porosity & $\mathrm{-}$\\
    $p$ & change in pressure & $\mathrm{Pa}$ \\
    $s$ & Laplace transform parameter & $\mathrm{-}$ \\
    $Q$ & volumetric flowrate & $\mathrm{m^3/s}$ \\
    $r$ & distance coordinate & $\mathrm{m}$ \\
    $r_w$ & borehole or excavation radius & $\mathrm{m}$ \\
    $\hat{\alpha}$ & \citet{warren1963behavior} shape factor & $\mathrm{1/m^2}$ \\
    $\eta$ & porosity power-law exponent & $\mathrm{-}$ \\ 
    $\kappa$ & permeability power-law exponent & $\mathrm{-}$ \\
    $\rho$ & fluid density & $\mathrm{kg/m^3}$ \\
    $\mu$ & fluid viscosity & $\mathrm{Pa \cdot s}$\\
    \hline
  \end{tabular}
  \caption{Physical Properties and Parameters}
  \label{tab:notation}
\end{table}

\begin{table}[htb]
  \centering
  \begin{tabular}{lll}
    \hline
    $p_D$ & scaled pressure & $p/p_c$ \\
    $t_D$ & scaled time & $t k_0/n_0 c L_c^2 \mu$ \\
    $r_D$ & scaled distance & $r/L_c$ \\
    $\lambda$ & interporosity exchange coefficient & $\hat{\alpha} k_r r_w^2/k_f$ \\
    $\sigma$ & wellbore storage coefficient & $A_c/(r_w n_0 c \rho g A_m)$ \\
    $\omega$ & fracture storage coefficient & $n_{f0}c_f/(n_{r0}c_r + n_{f0}c_f)$ \\
    \hline
  \end{tabular}
  \caption{Dimensionless Quantities}
  \label{tab:dimless}
\end{table}

\section*{Statements and Declarations}
\subsection*{Funding}
The author thanks the U.S. Department of Energy Office of Nuclear Energy's Spent Fuel and Waste Science and Technology program for funding.

\subsection*{Conflicts of Interest}
The author has no competing interests to declare.

\subsection*{Availability of Data and Material}
No data or materials were used by the author in the preparation of the manuscript.

\subsection*{Code Availability}
The source code of Fortran and Python implementations of the program are available from the author upon request.

\section*{Acknowledgments}
This paper describes objective technical results and analysis. Any subjective views or opinions that might be expressed in the paper do not necessarily represent the views of the U.S. Department of Energy or the United States Government.

This article has been authored by an employee of National Technology \& Engineering Solutions of Sandia, LLC under Contract No. DE-NA0003525 with the U.S. Department of Energy (DOE). The employee owns all right, title and interest in and to the article and is solely responsible for its contents. The United States Government retains and the publisher, by accepting the article for publication, acknowledges that the United States Government retains a non-exclusive, paid-up, irrevocable, world-wide license to publish or reproduce the published form of this article or allow others to do so, for United States Government purposes. The DOE will provide public access to these results of federally sponsored research in accordance with the DOE Public Access Plan \url{https://www.energy.gov/downloads/doe-public-access-plan}.

The author thanks Tara LaForce from Sandia for technically reviewing the manuscript.

\section{Appendix A: Wellbore Storage Boundary Condition}
\label{sec:WBIIIderivation}
The wellbore-storage boundary condition accounts for the storage in the finite borehole arising from the mass balance $Q_{\mathrm{in}}-Q_{\mathrm{out}}=A_c \frac{\partial h_w}{\partial t}$. $Q_{\mathrm{in}}$ $\mathrm{[m^3/s]}$ is volumetric flow into the borehole from the formation, $Q_{\mathrm{out}}$ is possibly time-variable flow out of the well through the pump ($Q(t)$ $\mathrm{[m^3/s]}$), and $\frac{\partial h_w}{\partial t}$ is the change in hydraulic head [m] ($h_w=\frac{p_w}{\rho g} + z$) of water standing in the borehole through time, $p_w$ is change in pressure [Pa] of water in the borehole, $\rho$ is fluid density $\mathrm{[kg/m^3]}$, $z$ is an elevation datum [m], and $g$ is gravitational acceleration $\mathrm{[m/s^2]}$. $A_c$ is the cross-sectional surface area of the pipe, sphere or box providing storage (it may be a constant or a function of elevation); for a typical pipe, it becomes $A_c=\pi r_c^2$, where $r_c$ is the radius of the casing where the water level is changing.  The mass balance is then 
\begin{equation}
  \label{eq:III1}
  \frac{A_m k_0}{\mu} \left. \frac{\partial p}{\partial r} \right|_{r=r_w} - Q(t) = \frac{A_c}{\rho g} \frac{\partial p_w}{\partial t},
\end{equation}
where $A_m$ is the area of the borehole communicating with the formation.  For the integer $m$ considered here these are $A_0=b^2$, $A_1=2 \pi r_w b$, $A_2=4 \pi r_w^2$  ($b$ is a length independent of the borehole radius). 

Assuming the change in water level in the borehole ($h_w=p_w/\left(\rho g\right)$) is equal to the change in formation water level ($h=p/\left(\rho g\right)$), this can be converted into dimensionless form as
\begin{equation}
  \label{eq:III2}
  \left. \frac{\partial p_D}{\partial r_D} \right|_{r_D=1} - f_t = \sigma \frac{\partial p_D}{\partial t},
\end{equation}
where $\sigma=A_c/\left( r_w n_0 c \rho g A_m \right)$ is a dimensionless ratio of formation to wellbore storage; $\sigma \rightarrow 0$ is an infinitesimally small well with only formation response, while $\sigma \rightarrow \infty$ is a well with no formation response (i.e., a bathtub).

\section{Appendix B: Transformation of Modified Bessel Equation}
\label{sec:BEderivation}
Following the approach of \citet{bowman58}, alternative forms of the Bessel equation are found, this approach is a simplification of the original approach of \citet{lommel68}. An analogous approach is applied here to ``back into'' the desired modified Bessel equation. The equation satisfied by the pair of functions
\begin{align}
  \label{eq:app1}
  y_1&=x^\alpha \mathrm{I}_\nu \left( \beta x^\gamma \right), & y_2&=x^\alpha \mathrm{K}_\nu \left( \beta x^\gamma \right)
\end{align}
is sought, where $\alpha$, $\beta$, $\gamma$, and $\nu$ are constants. Using the substitutions $\zeta=yx^{-\alpha}$ and $\xi=\beta x^\gamma$ gives $\zeta_1=\mathrm{I}_\nu\left( \xi \right)$ and $\zeta_2=\mathrm{K}_\nu\left( \xi \right)$, which are  the two solutions to the modified Bessel equation \citep[\S 10.25.1]{NIST:DLMF},
\begin{equation}
  \label{eq:app4}
  \xi \frac{\mathrm{d}}{\mathrm{d}\xi} \left( \xi \frac{\mathrm{d}\zeta}{\mathrm{d}\xi}\right) - (\xi^2 + \nu) \zeta = 0.
\end{equation}
Given
\begin{equation}
  \label{eq:app6}
  \xi \frac{\mathrm{d}}{\mathrm{d}\xi} \left( \xi \frac{\mathrm{d}\zeta}{\mathrm{d}\xi}\right) = \frac{x}{\gamma^2} \frac{\mathrm{d}}{\mathrm{d}x} \left( x \frac{\mathrm{d}\zeta}{\mathrm{d}x} \right),
\end{equation}
and
\begin{equation}
  \label{eq:app8}
x \frac{\mathrm{d}}{\mathrm{d}x} \left( x \frac{\mathrm{d}\zeta}{\mathrm{d}x} \right) = \frac{y''}{x^{\alpha-2}} - \frac{\left(2 \alpha-1 \right)y'}{x^{\alpha-1}} + \frac{\alpha^2 y}{x^\alpha},
\end{equation}
the standard-form equation satisfied by $y$ is
\begin{equation}
  \label{eq:app9}
  y'' + \left(1-2 \alpha \right)y' + \frac{\alpha^2 y}{x^\alpha} - \left( \beta^2 \gamma^2 x^{2\gamma-2}  - \frac{\alpha^2 - \nu^2 \gamma^2}{x^2} \right)y = 0.
\end{equation}
This equation can be compared to the Laplace-space ordinary differential equation \eqref{eq:non-dim-lap-gen}, allowing direct use of the product of powers and modified Bessel function \eqref{eq:app1} as solutions \eqref{eq:lommel-soln}.

\bibliography{transport}

\end{document}